\documentclass[12pt]{article}
\usepackage{amsmath}
\newcommand{\be}{\begin{equation}}\newcommand{\ee}{\end{equation}}
\newcommand{\bea}{\begin{eqnarray}}\newcommand{\eea}{\end{eqnarray}}
\newcommand{\ba}{\begin{array}}\newcommand{\ea}{\end{array}}
\newcommand{\p}[1]{(\ref{#1})}
\newcommand{\nn}{\nonumber}

\newcommand{\deriv}[2]{\frac{\partial #1}{\partial #2}\,}

\newcommand{\bD}{{\overline D}}
\newcommand{\bx}{{\bar x}}
\newcommand{\by}{{\bar y}}
\newcommand{\bxi}{{\bar\xi}}
\newcommand{\bpsi}{{\bar\psi}}
\newcommand{\blam}{{\bar\lambda}}

\begin{document}

\thispagestyle{empty}
\vspace{2cm}
\begin{flushright}
%Third draft \\
%15.10.2007\\
CERN-PH-TH/2007-194\\
ITP--UH--23/07
\end{flushright}\vspace{2cm}
\begin{center}
{\Large\bf Superfield Formulation of \\[12pt]
 Nonlinear N=4 Supermultiplets}
\end{center}
\vspace{1cm}

\begin{center}
{\large\bf S.~Bellucci${}^{a}$, S.~Krivonos${}^{b}$,
O.~Lechtenfeld${}^{c}$, A.~Shcherbakov${}^{a,b}$ }
\end{center}

\begin{center}
${}^a$ {\it
INFN-Laboratori Nazionali di Frascati,
Via E. Fermi 40, 00044 Frascati, Italy} \vspace{0.2cm}

${}^b$ {\it
Bogoliubov  Laboratory of Theoretical Physics, JINR,
141980 Dubna, Russia} \vspace{0.2cm}

${}^c$ {\it
Theory Division, Physics Department, CERN,
1211 Geneva 23, Switzerland} \\
{\it on leave of absence from:
Leibniz Universit\"at Hannover, Germany} \vspace{1cm}

\end{center}
\vspace{2cm}

\begin{abstract}\noindent
We propose a unified superfield formulation of $N{=}4$ off-shell
supermultiplets in one spacetime dimension using the standard
$N{=}4$ superspace. The main idea of our approach is a ``gluing''
together of two linear supermultiplets along their fermions.
The functions defining such a gluing obey a system of equations.
Each solution of this system provides a new supermultiplet,
linear or nonlinear, modulo equivalence transformations.
In such a way we reproduce all known linear and nonlinear
$N{=}4$, $d{=}1$ supermultiplets and propose some new ones.
Particularly interesting is an explicit construction of nonlinear
$N{=}4$ hypermultiplets.
\end{abstract}

\newpage
\setcounter{page}{1}
\section{Introduction}
The main ingredients for the construction of one-dimensional systems
with extended $N\geq 4$ supersymmetry are irreducible supermultiplets.
Given a set of those, preferably formulated in superspace,
one may immediately write the corresponding sigma-model type actions
and the  general potential terms.
In this respect, the almost complete classification of {\it linear\/} off-shell
representations for one-dimensional supersymmetry \cite{GR,JG,FT} seems
to suffice for constructing any mechanics model with extended supersymmetry.
However, detailed analysis of the corresponding actions reveals a common
restriction -- the bosonic parts of all actions describe only conformally
flat manifolds. Moreover, the prepotentials describing the most general
interaction are constrained to obey flat Laplace equations in superspace.
These are signals that something essential is missing. The above-mentioned
classification of linear representations admits only one possibility:
there must exist additional {\it nonlinear\/} representations.

The possibility of nonlinear off-shell $N{=}4$, $d{=}1$ supermultiplets
was firstly noted in~\cite{il1}.
Subsequently, in~\cite{nlchiral} the first two examples of such nonlinear
supermultiplets were explicitly described.
One of these examples was reduced from a four-dimensional cousin while
the other one was completely new.
The next step was taken in~\cite{delduc1}, with the reduction of the $N{=}2$,
$d{=}4$ hypermultiplet to an off-shell $N{=}4$, $d{=}1$ supermultiplet.
These new nonlinear supermultiplets with four physical bosonic and
four fermionic components were explicitly constructed~\cite{ks1},
and their formulation in harmonic superspace was proposed~\cite{delduc2}.
In parallel, the component description of several new nonlinear $N{=}8$
supermultiplets was found~\cite{N8}.

Although by now the list of nonlinear $N{=}4$ supermultiplets has
gotten a bit lengthy, no attempt has as yet been made for their
classification. The main obstacle here is the variety of methods
by which these supermultiplets have been constructed: Some have
been found within the geometric approach based on a nonlinear
realization of the $N{=}4$ superconformal group~\cite{nlchiral},
others were built by applying the so-called dualization
procedure~\cite{ks1}. In further cases, the harmonic superspace
constraints just mimic their $N{=}4$, $d{=}2$
counterparts~\cite{il1,delduc2}. Moreover, part of these nonlinear
supermultiplets have been formulated in terms of components, part
in the standard $N{=}4$, $d{=}1$ superspace, while the rest in
harmonic superspace. Clearly, for a classification it is desirable
to have a unified description. Yet, such a framework has to be
flexible enough not to exclude nonlinear supermultiplets which
have yet to be discovered.

The main goal of the present paper is to provide such a unified approach
towards nonlinear supermultiplets with $N{=}4$ supersymmetry in one spacetime
dimension.
The key idea is to construct a nonlinear supermultiplet by entangling a pair
of linear $N{=}4$ supermultiplets.
Let us illustrate the main steps of our construction.

For the sake of clarity we momentarily suppress all indices but one.
A linear $N{=}4$ supermultiplets consists of $n$~physical fields~$\phi$,
$4$~fermionic ones~$\psi$, and $4{-}n$~auxiliary ones~$A$.
Taking two such supermultiplets $\Phi_1$ and~$\Phi_2$
with $n_1$ and $n_2$~physical bosons $\phi_1$ and~$\phi_2$, respectively,
we have twice as many fermions $\psi_1$ and~$\psi_2$ as is
required by $N{=}4$ supersymmetry. This is not a problem in principle,
but to get the minimal representation we must reduce this amount
by somehow identifying the fermions of both supermultiplets.
Denoting by $D$ and~$\bar D$ the covariant spinor derivatives, so that
$D \Phi_{1,2}=\psi_{1,2}+\ldots$,
the most general identification of the two sets of four spinors reads
\be\label{eq0}
D\Phi_1\ =\ f D\Phi_2 + g\bar D\Phi_2 + h D\bar\Phi_2 + k\bar D\bar\Phi_2
\qquad\textrm{with functions $f$, $g$, $h$, $k$ of $\Phi_1$ and $\Phi_2$}\ .
\ee
As a consequence, the resulting {\it nonlinear\/} representation contains only four independent fermions rather than eight. Due to
supersymmetry, some of the higher components of~$\Phi_{1,2}$ will be expressed through lower components. The total number of physical
components of the combined representation is just $n_1{+}n_2$, leaving $4-(n_1{+}n_2)$ auxiliary fields in total. Since all numbers must be
non-negative, the possibilities are restricted by the inequality~$n_1+n_2\leq4$. It turns out that a vanishing $n_1$ or $n_2$ will just
reproduce the partner supermultiplet, and so the nontrivial list of cases is
\be\label{list}
(n_1,n_2)\ =\ (1,1)\ ,\quad (2,1)\ ,\quad (3,1)\ ,\quad (2,2)\ .
\ee

Clearly enough, the functions $f$, $g$, $h$ and $k$ cannot be
completely arbitrary, because the irreducible $N{=}4$ superfields
$\Phi_{1,2}$ obey some constraints. These constraints imply a
system of equations which these functions must satisfy. Each
solution to this system gives rise to some irreducible
supermultiplet (linear or nonlinear). Of course, some of these
solutions may be equivalent to others via some superfield
redefinition. Modulo this freedom, one may expect to find some of
the known supermultiplets among the solutions. However, it is
unexpected -- and very satisfying -- to see that actually {\it all
known\/} linear and nonlinear supermultiplets may be constructed
in this fashion. Moreover, the set of solutions is large enough to
leave room for yet undiscovered nonlinear supermultiplets. The
next four sections are devoted to the derivation of these results
in the cases~(\ref{list}). A final section comments upon the
implications of the results for the classification problem and
touches upon a number of related issues.

\setcounter{equation}0
\section{2=1+1}
Let us start with the simplest example. Our goal is to construct
the irreducible $N{=}4, d{=}1$ supermultiplet with the two
physical bosons starting from two irreducible $N{=}4$
supermultiplets containing one physical boson each. In terms of
the $N{=}4$ superfields the $N{=}4$ supermultiplet with one
physical boson is completely defined by a scalar superfield
obeying the constraints \cite{leva}. We need two such
supermultiplets, so we introduce two scalar $N{=}4$ superfields
$u$ and $v$ satisfying \be\label{143} D^{(i}\bD{}^{j)} u=0, \quad
D^{(i}\bD{}^{j)} v=0. \ee Here, $D$ and $\bD$ are spinor covariant
derivatives obeying the standard super Poincar\'e algebra \be
\left\{ D^i, \bD_j \right\} = 2i \delta^i_j \partial_t, \ee and
the brackets $()$ as usually mean symmetrization over the indices
enclosed.

Each of our superfields $u,v$ contains among the components one
physical scalar, four physical fermions and three auxiliary
bosons. Clearly, to get the irreducible supermultiplet one has
reduce the number of the physical fermions to four. The simplest
way to do this is to identify the spinors in both supermultiplets.
The most general identifications is achieved in the following way:
\be\label{eq143} D^i u = f_1 D^i v + f_2 \bD^i v, \qquad \bD_i u =
{\bar f}_1 \bD_i v - {\bar f}_2 D_i v, \ee with the arbitrary
functions~$f_1, f_2$ depending on both superfields~$u$ and~$v$.

The superfield $u$ obeys~\p{143}, therefore the r.h.s in the \p{eq143} should be also antisymmetric over $su(2)$ indices upon action of
$D^j$ and $\bD{}^j$ on them. This leads to the constraints on the functions~$f_{1,2}$
\bea\label{constr143}
&&f_2 \deriv{f_1}u = f_1 \deriv{f_2}u + \deriv{f_2}v, \qquad -\bar f_2 \deriv{f_2}u = \bar f_1 \deriv{f_1}u +
\deriv{f_1}v, \nn\\
&&\bar f_2 \deriv{\bar f_1}u = \bar f_1 \deriv{\bar f_2}u +
\deriv{\bar f_2}v, \qquad -f_2 \deriv{\bar f_2}u = f_1 \deriv{\bar
f_1}u + \deriv{\bar f_1}v. \eea Let us note that the equations
\p{eq143}, being satisfied, reduce also the number of the
auxiliary components to two in both supermultiplets expressing
some of the auxiliary components through time derivatives of the
physical bosons and identifying the remaining ones in both
supermultiplets. Thus any solution of the system \p{constr143}
provides us with the $N{=}4$ supermultiplet with two physical
bosons.

Before going on to solve the equations \p{constr143}, one should
note that we are free to choose the basic superfields in a
different way. Indeed, one may write, for example, the basic
constraint on the general superfunction $G(u,v)$ depending in an
arbitrary way on $u$ and $v$ \be \left\{ \begin{array}{l}
D^i G = f_1 D^i v + f_2 \bD^i v, \\
\bD_i G = {\bar f}_1 \bD_i v - {\bar f}_2 D_i v,
\end{array} \right. \Rightarrow
\left\{ \begin{array}{l}
D^i u = \frac{ f_1-\deriv{G}v}{\deriv{G}u} D^i v + \frac{f_2}{\deriv{G}u} \bD^i v, \\
\bD^i u = \frac{ {\bar f}_1-\deriv{G}v}{\deriv{G}u} D^i v - \frac{{\bar f}_2}{\deriv{G}u} D_i v.
\end{array} \right.
\ee Clearly, one may use this gauge freedom to completely remove
the real part of the function $f_1$. Thus, from now we  impose the
following condition: \be\label{sol143a} f_1= if, \quad {\bar
f}_1=-if, \ee where $f(u,v)$ is a real function.

Now we are ready to find the general solution of the equations
\p{constr143}. First of all, one may easily show that from the
equations \p{constr143} and \p{sol143a} it follows that
\be\label{sol143b} \frac{\partial}{\partial u}\left( f^2+ f_2
{\bar f}_2 \right)=0,\quad \frac{\partial}{\partial v}\left( f^2+
f_2 {\bar f}_2 \right)=0. \ee Therefore, $ f^2+ f_2 {\bar
f}_2=const$ and we are free to fix this constant
\be\label{sol143c}
 f^2+ f_2 {\bar f}_2 =1.
\ee Now, it is rather convenient to solve the equation \p{sol143c}
as \be\label{sol143d} f=\frac{{h\bar h}-1}{{h\bar h}+1},\qquad
f_2=\frac{2ih}{h{\bar h}+1}, \qquad {\bar f}_2=-\frac{2i{\bar
h}}{{h\bar h}+1}, \ee where $h,{\bar h}$ are two arbitrary
functions. Substituting \p{sol143d} in \p{constr143} we will get
the following equations: \be\label{sol143e} {\bar h}\left( ih_u
-h_v\right)=0,\qquad h\left(i{\bar h}_u+{\bar h}_v\right)=0 \ee
which have the evident solution \be\label{sol143f}
h=h(u+iv),\qquad {\bar h}={\bar h}(u-iv). \ee Thus, our basic
constraints \p{constr143} read \be\label{constr143a} D^iu
=i\frac{h{\bar h}-1}{h{\bar h}+1} D^i v+\frac{2ih}{h{\bar
h}+1}\bD{}^iv,\quad \bD^iu =-i\frac{h{\bar h}-1}{h{\bar h}+1}
\bD^i v+\frac{2i{\bar h}}{h{\bar h}+1}D_iv. \ee The last step is
to rewrite the system \p{constr143a} as \be\label{sol143g}
D^i\left(u+iv\right)=h(u+iv)\bD{}^i\left(u+iv\right), \quad
\bD{}^i\left(u-iv\right)=-{\bar h}(u+iv)D^i\left(u-iv\right). \ee
So, one may construct the $N{=}4$ supermultiplet with the two
physical bosons  from the two supermultiplets with one physical
bosons by imposing on them the constraints \p{sol143g}.

It is quite easy to recognize which supermultiplets we constructed. If the functions $h={\bar h}=0$,
the constraints \p{sol143g} describe the standard $N{=}4$ chiral supermultiplet. If the function $h=const$,
then we deal with the twisted chiral supermultiplet. Finally, if $h\neq const$ one may multiply the equations
in \p{sol143g} by $h'$ and ${\bar h}{}'$ respectively, to get
\be\label{sol143}
D^i Z= Z\bD{}^i Z, \quad \bD_i {\bar Z}=-{\bar Z} D_i {\bar Z},
\ee
where
\be
Z\equiv h(u+iv),\quad {\bar Z} \equiv {\bar h}(u-iv).
\ee
The constraints \p{sol143} defined the nonlinear chiral supermultiplet \cite{nlchiral}.

Thus,  we were able to construct all known $N{=}4, d{=}1$ supermultiplets with the two
physical bosons among the components. Moreover, no other solutions exist within our approach. This is in a full
agreement with the claim of the paper \cite{di1} that all possible two dimensional supermultiplets
include chiral and nonlinear chiral supermultiplets only.

\setcounter{equation}{0}
\section{3=2+1}
In this Section we will construct the $N{=}4$ supermultiplets with
three physical bosons starting from two supermultiplets with one
and two physical bosons, respectively. To describe the $N{=}4$
supermultiplet with one physical boson we will use the same real
$N{=}4$ superfield $u$ as in the previous Section, subjected to
the constraints \p{143}. In addition, the chiral $N{=}4$
superfield $\lambda,\blam$ \be\label{242} D^i \lambda=0, \quad
\bD{}^i \blam=0, \ee contains just two physical boson components.
Now we have to identify the fermionic components in both
supermultiplets as \be\label{con242} D^i u = f_1 D^i \blam +f_2
\bD{}^i \lambda, \quad \bD{}^i u = {\bar f}_1 \bD{}^i \lambda-
{\bar f}_2 D^i \blam, \ee where $f_{1,2}(u,\lambda, \blam)$ are
arbitrary functions depending on all our superfields
$(u,\lambda,\blam)$.

As well as in the previous case, the consistency of \p{con242}
imposes the restrictions on the functions $f_{1,2}$ \bea && f_2
\frac{\partial f_1}{\partial u}=f_1 \frac{\partial f_2}{\partial
u}+ \frac{\partial f_2}{\partial \blam}, \quad (a) \qquad {\bar
f}_2 \frac{\partial{\bar f}_1}{\partial u}={\bar f}_1
\frac{\partial {\bar f}_2}{\partial u}+ \frac{\partial
{\bar f}_2}{\partial \lambda}, \quad (b) \label{eq2421}\\
&&-{\bar f_2}\frac{\partial f_2}{\partial u}={\bar
f}_1\frac{\partial f_1}{\partial u}+\frac{\partial f_1}{\partial
\lambda}, \quad (a) \qquad -{f_2}\frac{\partial {\bar
f}_2}{\partial u}={f}_1\frac{\partial {\bar f}_1}{\partial
u}+\frac{\partial {\bar f}_1}{\partial \blam} \quad (b).
\label{eq2422} \eea So, any solution of the systems \p{eq2421},
\p{eq2422} describes the irreducible $N{=}4$ supermultiplet with
three physical bosons, modulo possible redefinitions of the
superfields. To partially fix this freedom, let us note that we
may write the same equations \p{eq2421}, \p{eq2422} on the
arbitrary real superfunction $G(u,\lambda,\blam)$ instead of $u$.
This will result in the same equations \p{eq2421}, \p{eq2422} for
the superfield $u$ but with the modified functions ${\tilde
f}_{1,2}$ \be\label{freedom2} {\tilde f}_1=\frac{1}{G_u}\left(
f_1-G_{\blam}\right), \quad {\tilde{\bar
f}}_1=\frac{1}{G_u}\left({\bar f}_1-G_\lambda\right), \qquad
{\tilde f}_2=\frac{1}{G_u}f_2, \; {\tilde {\bar
f}}_2=\frac{1}{G_u}{\bar f}_2. \ee Using this freedom we cannot
fully remove the real part of the function $f_1$ as in the
previous Section. Instead, one may partially restrict $f_1$
imposing the following condition: \be\label{eq242c} \frac{\partial
}{\partial \lambda} f_1 +\frac{\partial}{\partial \blam}{\bar
f_1}=0. \ee

Before solving the systems \p{eq2421}, \p{eq2422}, \p{eq242c} it is useful to demonstrate how the known
$N{=}4$ supermultiplets with three physical bosons appear among the solutions of these equations.

\subsection{Linear tensor supermultiplet}
The linear tensor supermultiplet \cite{tensor} is defined in terms
of the $su(2)$ triplets of the bosonic superfields $V^{(ij)}$
subjected to the following constraints: \be\label{lintensor}
\nabla^{(i}V^{jk)}=0, \quad \bar\nabla{}^{(i}V^{jk)}=0, \ee where
$\nabla^i, \bar\nabla{}^i$ is the set of $N{=}4$ covariant
derivatives with the standard superalgebra \be \left\{ \nabla^i,
{\bar\nabla}_j \right\} = 2i \delta^i_j \partial_t. \ee Redefining
the superfields and the covariant derivatives as \bea\label{newf}
&& V^{ii}=\lambda,\quad V^{22}=\blam, \quad V^{12}=iu, \nn\\
&& D^1=\nabla^1,\; D^2=-\bar\nabla_2,\quad \bD_1= \bar\nabla_1, \; \bD_2=-\nabla^2,
\eea
one may rewrite the basic constraints \p{lintensor} as
\bea\label{lin242}
&& D^i \lambda=0,\quad \bD_i \blam=0, \nn \\
&& D^i u= \frac{i}{2} \bD{}^i \lambda, \quad \bD^i u =\frac{i}{2} D^i \blam.
\eea
Clearly, the constraints \p{lin242} coincide with \p{con242} if we choose
\be\label{linsol}
f_1={\bar f}_1=0, \qquad f_2=\frac{i}{2}, \quad {\bar f}_2=-\frac{i}{2}.
\ee
It is trivial to check that \p{linsol} is a particular solution of the system \p{eq2421}, \p{eq2422}, \p{eq242c}.

\subsection{Nonlinear tensor supermultiplet}
The nonlinear tensor supermultiplet is defined in terms of the three bosonic $N{=}4$ superfields $(u,\lambda,\blam)$ obeying the
constraints \cite{il1,nlchiral}
\bea\label{nltensor}
&& D^i \lambda=0, \quad \bD{}^i \blam=0, \nn \\
&& D^i \left( e^{-iu}\blam\right) =-i \bD{}^i u, \quad \bD{}^i \left( e^{iu}\lambda\right)=-i D^i u.
\eea
Rewriting the second line in the system \p{nltensor} as
\be\label{nltensor1}
D^i u = \frac{i}{1+\lambda\blam}\left[ -\lambda D^i \blam +e^{iu} \bD{}^i \lambda\right], \;
\bD^i u = \frac{i}{1+\lambda\blam}\left[ \blam \bD{}^i \lambda +e^{-iu} D^i \blam\right],
\ee
one may again find the full agreement with \p{con242} upon identification
\be\label{nlinsol}
f_1=-i\frac{\lambda}{1+\lambda\blam}, \quad f_2 = i \frac{e^{iu}}{1+\lambda\blam}.
\ee
The expressions \p{nlinsol}, like to the previous case, provide the particular solution of the
equations \p{eq2421}, \p{eq2422}, \p{eq242c}.

\subsection{General solution}
Thus, all known supermultiplets with the three physical bosons are present among the solutions of our system \p{eq2421}, \p{eq2422},
\p{eq242c}. To understand whether there are other solutions describing new $N=4$ supermultiplets one has to solve the equations \p{eq2421},
\p{eq2422}, \p{eq242c}.

First of all, let us note that the superfields $\lambda,\blam$ and
the covariant derivatives are charged with respect to $U(1)$
rotations \be\label{u1} D^i \rightarrow e^{i\alpha} D^i, \; \bD^i
\rightarrow e^{-i\alpha} \bD^i,\quad \lambda \rightarrow
e^{2i\alpha} \lambda,\; \blam \rightarrow e^{-2i\alpha} \blam, \ee
while the superfield $u$ is chargeless. To keep this $U(1)$
invariance manifest, let us suppose that our functions $f_{1,2}$
are restricted as \be\label{restr1} f_1=\lambda {\tilde f}_1(u,
z),\quad f_2=f_2(u,z),\qquad z \equiv \lambda\blam. \ee With these
conditions, the equation \p{eq242c} reads \be\label{eq2a}
\frac{\partial}{\partial z} \left[ z \left( {\tilde f}_1+
\bar{\tilde f}_1\right)\right]=0. \ee Thus, the real part of the
function ${\tilde f}_1$ is completely fixed to be \be\label{sol2a}
{\tilde f}_1+ \bar{\tilde
f}_1=\frac{F(u)}{z}\equiv\frac{F(u)}{\lambda\blam}, \ee where
$F(u)$ is an arbitrary function depending on the superfield $u$
alone. Substituting \p{sol2a} into our basic constraints
\p{con242} one may easily check that one can always redefine the
superfield $u$ to cancel this part in the constraints. So, from
now on, we will impose the further restriction on the functions
${\tilde f}_1$ \be\label{sol2b} {\tilde f}_1=if(u,z), \qquad
\bar{\tilde f}_1=-if(u,z). \ee Thus, the equation \p{eq242c} is
satisfied, while the systems \p{eq2421}, \p{eq2422} read \bea && i
f_2 \frac{\partial f}{\partial u}=i f \frac{\partial f_2}{\partial
u}+\frac{\partial f_2}{\partial z}\quad (a), \qquad
-i{\bar f}_2 \frac{\partial f}{\partial u}=-i f \frac{\partial {\bar f}_2}{\partial u}+\frac{\partial {\bar f}_2}{\partial z}\quad (b), \label{242a} \\
&& z f\frac{\partial f}{\partial u} +i\frac{\partial}{\partial
z}\left( z f\right)=-{\bar f}_2\frac{\partial f_2}{\partial u}
\quad (a), \qquad z f\frac{\partial f}{\partial u}
-i\frac{\partial}{\partial z}\left( z
f\right)=-{f}_2\frac{\partial {\bar f}_2}{\partial u} \quad (a).
\label{242b} \eea Summing (\ref{242b}a) and (\ref{242b}b) we will
get the equation \be\label{242e} \frac{\partial}{\partial u}
\left[ z f^2 +f_2{\bar f_2}\right]=0, \ee while the difference of
these equations produces \be\label{x1} 2i\frac{\partial}{\partial
z} \left( z f\right)=f_2 \frac{\partial {\bar f}_2}{\partial u} -
{\bar f}_2 \frac{\partial {f}_2}{\partial u}. \ee If we further
sum the equation (\ref{242a}a) multiplied by ${\bar f}_2$ with the
equation (\ref{242a}b) multiplied by $f_2$ we will obtain the
equation \be\label{x2} \frac{\partial}{\partial z}\left( f_2{\bar
f}_2\right) =if \left[ f_2\frac{\partial {\bar f}_2}{\partial u} -
{\bar f}_2 \frac{\partial {f}_2}{\partial u}\right]. \ee Now,
combining \p{x1} and \p{x2} we will have \be\label{x3}
\frac{\partial}{\partial z}\left[z f^2 +f_2{\bar
f}_2\right]+f^2=0. \ee In virtue of \p{242e} we immediately
conclude from \p{x3} that \be\label{sol3} \frac{\partial}{\partial
u} f =0, \ee and therefore \be\label{sol4}
\frac{\partial}{\partial u} \left( f_2{\bar f}_2\right)=0 \quad
\Rightarrow f_2=h(z)e^{i\Psi(z,u)},\; {\bar f}_2={\bar h}(z)
e^{-i\Psi(z,u)}. \ee Moreover, plugging \p{sol4} in the equation
\p{x1} one may find that $\frac{\partial}{\partial u} \Psi(z,u)$
does not depend on $u$, and therefore $\Psi(z,u)=\alpha u,
\alpha=const$, modulo redefinitions of $h, {\bar h}$. Finally, it
follows from \p{242b} that \be h' {\bar h} - h {\bar h}{}'=0 \quad
\Rightarrow h=\beta e^{\Phi(z)}, \; {\bar h}=\bar\beta
e^{\Phi(z)},\quad \beta=const. \ee

Putting all these together, we have the following semi-solution of our basic system \p{eq2421}, \p{eq2422}, \p{eq242c}
\be\label{x4}
f_1=i\lambda f(z), \; {\bar f}_1=-i\blam f(z),\quad f_2=\beta e^{i\alpha u+\Phi(z)},\;
{\bar f}_2=\bar\beta e^{-i\alpha u+\Phi(z)},
\ee
where two real functions $f(z)$ and $\Phi(z)$ are still restricted to obey
\be\label{x5}
\frac{d}{dz}\left(z f\right)=-\alpha \beta\bar\beta e^{2\Phi},\quad \frac{d}{dz}\Phi=\alpha f.
\ee
If $\alpha=0$ then the solution of \p{x5} is trivial and describes the linear tensor supermultiplet. Alternatively, with $\alpha\neq 0$ one
may always rescale the superfield $u$ to fix $\alpha=1$.

The general solution of the system \p{x5} with $\alpha=1$ reads
\be\label{gsol242}
f=\frac{-1+c_1 \left( -1 +\frac{2 c_2}{z^{c_1}+c_2}\right)}{2z}, \quad
e^\Phi=\frac{c_1 \sqrt{c_2}\; z^{\frac{1}{2}\left( c_1 -1\right)}}{\sqrt{\beta\bar\beta}\left( z^{c_1}+c_2\right)},
\ee
where $c_1, c_2$ are arbitrary real constants.

\subsection{Action}
To understand better what systems can be described  by the new nonlinear supermultiplet let us construct the action. The general sigma
model type action may be easily constructed as the integral over $N{=}4$ superspace
\be\label{action1a}
S_1= \int dt d^2 \theta d^2 \bar\theta \; L(u,\lambda, \blam).
\ee
Here, $L(u,\lambda, \blam)$ is an arbitrary real function.

Before going to the component action and to possible potential terms one has to understand the structure of the auxiliary bosonic
components in our supermultiplet. From the beginning we have three auxiliary components in the superfield $u$ and two components in the
superfield $\lambda,\blam$ :
\be\label{aux1}
A= D^i D_i u|, \; C=[D^i,\bD_i]\;u|, \; {\bar A}=\bD{}^i\bD_i u|, \qquad
B=\bD{}^i\bD_i \lambda|, \; {\bar B}=D^i D_i \blam|,
\ee
where $|$ means limit $\theta=\bar\theta=0$. Let us concentrate on pure bosonic equations discarding all the fermionic terms. Thus, from
our basic constraints \p{con242} it immediately follows the relation between auxiliary components and time derivatives from physical
bosons:
\bea\label{aux2}
&& A=f_1 B+4if_2{\dot\lambda},\; {\bar A}={\bar f}_1 {\bar B}+4i{\bar f}_2{\dot\blam},\quad C=4i\left( {\bar f}_1 \dot\lambda -f_1\dot\blam\right)+f_2 B-{\bar f}_2{\bar B}, \nn \\
&&4i {\dot u}=4i\left( {\bar f}_1 \dot\lambda +f_1\dot\blam\right)-f_2 B-{\bar f}_2{\bar B}.
\eea
First of all, we conclude that the function $f_2$ can not be equal zero, because otherwise from \p{aux2} it follows the relation between
time derivatives of the physical bosonic components, and therefore we get the on-shell multiplet. With $f_2 \neq 0$ our constraints leave
only one auxiliary component in the superfields $u, \lambda, \blam$ as it should be.

Now one may construct the general potential term for our supermultiplet.  To do this, one should notice that from the constraints
\p{con242} it follows that all the spinor derivatives with respect to $\theta_2, \bar\theta{}^2$ may be expressed as $\theta_1$-- and
$\bar\theta{}^1$--derivatives:
\bea\label{t2}
&& D^2 \blam =\frac{{\bar f}_1 \bD_1 \lambda -\bD_1 u}{{\bar f}_2}, \quad
{\bD}_2 \lambda= \frac{f_1 D^1 \blam - D^1 u}{f_2}, \nn \\
&&D^2 u=\frac{\left( f_1{\bar f}_1+f_2{\bar f}_2\right)
\bD_1\lambda-f_1 \bD_1 u}{{\bar f}_2},\; \bD_2 u=\frac{\left(
f_1{\bar f}_1+f_2{\bar f}_2\right) D^1\blam-{\bar f}_1 D^1
u}{{f}_2}. \eea Thus, all the components are sitting in the
$N{=}2$ superfields $({\tilde u},\tilde\lambda,\tilde\blam)$ \be
\tilde u = u|_{\theta_2=\bar\theta{}^2=0},\; \tilde \lambda =
\lambda|_{\theta_2=\bar\theta{}^2=0},\; \tilde \blam =
\blam|_{\theta_2=\bar\theta{}^2=0}. \ee Therefore, the most
general potential term can be written as \be\label{pot1} S_2 =m
\int dt d\theta_1 d\bar\theta{}^1 F({\tilde
u},\tilde\lambda,\tilde\blam). \ee By construction, the potential
term \p{pot1} is invariant under $N{=}2$ supersymmetry realized on
the $(\theta_1,\bar\theta{}^2)$. To be invariant under the other
implicit $N{=}2$ supersymmetry, the prepotential $F$ has to obey
the following equation: \be\label{t3} \left( f_1{\bar f}_1
+f_2{\bar f}_2\right) F_{{\tilde u}{\tilde u}}+F_{\tilde\lambda
\tilde\blam}+ {\bar f}_1 F_{{\tilde u}\tilde\blam} +f_1 F_{{\tilde
u}\tilde\lambda}=0. \ee So, the most general action for our
nonlinear supermultiplet reads \be\label{action242} S=S_1+S_2=\int
dt d^2 \theta d^2 \bar\theta \; L(u,\lambda, \blam)+m \int dt
d\theta_1 d\bar\theta{}^1 F({\tilde u},\tilde\lambda,\tilde\blam),
\ee where the prepotential $F$ is defined as the solution of the
equation \p{t3}.

Finally, let us present the bosonic sector of the action \p{action242}
\be\label{action242comp}
\begin{aligned}
S=\int dt \biggl[ & g\left( f_2{\bar f}_2 {\dot\lambda}{\dot\blam}+ \frac14
\left({\bar f}_1\dot\lambda+f_1\dot\blam- {\dot u}\right)^2\right) + \\
 + & i m \left[ \left(f_1 F_u +F_\blam\right)\dot\blam -\left({\bar f}_1
F_u +F_\lambda\right)\dot\lambda\right] - m^2 \frac{F_u^2}{g}\biggr],
\end{aligned}
\ee where \be\label{metric242} g=\frac{16}{f_2{\bar f}_2}\left(
\left(f_1{\bar f}_1+ f_2{\bar f}_2\right)L_{uu}+ {\bar f}_1
L_{u\blam}+f_1 L_{u\lambda}+L_{\lambda\blam}\right). \ee We
checked that with the $g=1$ the sigma model part of the action
\p{action242comp} describes a conformally flat (the Weyl tensor is
vanishing here) constant positive curvature three dimensional
manifold. Of course, to make any final conclusion about this model
one has to fully analyze all fermionic terms. We postpone this
analysis for the future.

\setcounter{equation}0
\section{4=3+1}
The first way to construct off-shell nonlinear $N{=}4$
supermultiplet with a four physical bosons is to start with two
$N{=}4$ supermultiplets containing three and one physical bosons,
respectively, and then identify the fermionic degrees of freedom
in both supermultiplets.

The $N{=}4$ supermultiplet with three physical bosons is well
known \cite{tensor}. It is called linear tensor supermultiplet and
may be described by a real $N{=}4$ superfield~ $v^{ij}$
$$v^{ij}=v^{ji},\qquad \left(v^{ij}\right)^\dagger=v_{ij},\qquad
i,j=1,2.$$ subject to  the constraints
\begin{equation}\label{V}
D^{(i} v^{jk)} = \bar D^{(i} v^{jk)}=0.
\end{equation}
The constraints (\ref{V}) leave in the tensor supermultiplet   just three physical bosons $v^{ij}$, four fermions $\xi^i, \bxi_i$ and one
auxiliary boson $A$
\begin{equation}\label{a1}
v^{ij} =v^{ij}|,\quad
 \xi^i = \frac13 D^j v^i_j|,\quad
 \bxi{}^i = -\frac13 \bD_j v^{ij}| ,\quad
 A = \frac i6 D^i \bD{}^j v_{ij}|,
\end{equation}
where, as usual, the symbol $|$ means restriction to
$\theta=\bar\theta=0$.

The second supermultiplet with one physical boson we need is an
``old tensor" supermultiplet \cite{leva}. This supermultiplet may
be described by a real superfield $u$ subjected to the following
constraints: \be\label{U} D^i D_i \;u = \bD^i\bD_i \;u=0. \ee It
comprises one physical boson $u$, once again four fermions
$\psi^i,\bpsi_i$ and a triplet of auxiliary components $A^{(ij)}$
\be\label{comp2} u=u|,\; \psi^i =D^i u|, \; \bpsi_i =\bD_i u|, \;
A^{(ij)} =\frac{i}{2} \left[ \bD^{(i}, D^{j)} \right] u|. \ee One
should stress that the constraints \p{U} describe just the same
multiplet with one physical boson we used in the previous
Sections. The twisted form of the constraints we are using now is
preferable for the following reasons. Within our approach we will
identify the fermions in both the supermultiplets $v^{ij}$ and
$u$. Clearly, this identification will reduce the number of the
auxiliary components in both the supermultiplets $A$ \p{a1} and
$A^{ij}$ \p{comp2} to zero, by expressing all these components in
terms of four physical components $v^{ij}$ and $u$. To be
manifestly invariant under the $SU(2)$ symmetry realized on the
doublet indices $(i,j)$ the three auxiliary components in the
superfield $u$ have to form a vector with respect to $SU(2)$. In
this case they may be expressed as time derivatives of $v^{ij}$
(plus fermionic terms with the same $SU(2)$ structure). Just this
structure of the auxiliary components is provided by the
constraints \p{comp2}.

Now we will identify the fermions in both supermultiplets by
imposing the following constraints: \be\label{con1} D^i u =
\frac{1}{3} f\; D_j v^{ij} - \frac{1}{3} a^{ij} D^k v_{kj}, \quad
\bD^i u = \frac{1}{3} \bar f \; \bD_j v^{ij} - \frac{1}{3} \bar
a^{ij} \bD^k v_{kj}, \ee where the
functions~$f(u,u)$,~$a^{ij}(u,v)$ and their conjugated ones depend
on both supermultiplets. In order to have equations~\p{con1}
consistent with~\p{V} and~\p{U}, these functions have to be real
and obey the following equations: \bea && 2f \frac{\partial
f}{\partial u}+ a_{ij}\frac{\partial a^{ij}}{\partial u}-
2 \frac{\partial a^{ij}}{\partial v^{ij}}=0 , \label{c1} \\
&& f \frac{\partial a_{ij}}{\partial u}-a_{ij}\frac{\partial f}{\partial u}+
2\frac{\partial f}{\partial v^{ij}} - \frac12 \left( a_{ik}\frac{\partial a^k_{j}}{\partial u}+
a_{jk}\frac{\partial a^k_{i}}{\partial u}\right) +
\left( \frac{\partial a^k_i}{\partial v^{kj}}+\frac{\partial a^k_j}{\partial v^{ki}}
\right)=0. \label{c2}
\eea
As we already explained before, in virtue of \p{c1},\p{c2} the auxiliary components of both
supermultiplets are expressed in terms of four physical bosons $u, v^{ij}$
\bea\label{aux}
&& A= \frac1f \left( \dot u + \frac12 a_{ij}{\dot v^{ij}} + fermions  \right), \nonumber \\
&& A^{ij} = f {\dot v^{ij}} + \frac{1}{2}\left( a^{ik} {\dot v}^j_k + a^{jk} {\dot v}^i_k \right)
    + a^{ij}A + fermions .
\eea
Thus, we indeed have a nonlinear supermultiplet with four physical bosonic and
four fermionic degrees of freedom.

Concerning the action, in~$N{=}4$ superspace it may be easily constructed in the standard way as
\be\label{action1}
S = \int d^4\theta dt \; L(u,v),
\ee
where Lagrangian $L$ is an arbitrary real function on superfields $u$ and $v^{ij}$.
Passing to the components one may easily find the bosonic part of the action \p{action1}
\be\label{action2}
S_{bos}=\int dt\; G \left[ \frac{1}{2} {\dot v^{ij}}{\dot v_{ij}}+
\frac{1}{f^2}\left( \dot u + \frac{1}{2} a_{ij}{\dot v^{ij}}\right)^2\right],
\ee
with the metric $G$
\be\label{metric1}
G\equiv \left( 2 f^2 + a_{ij}a^{ij}\right) \frac{\partial^2 L}{\partial u^2}-
4 a^{ij}\frac{\partial^2 L}{\partial u \partial v^{ij}}+
4 \frac{\partial^2 L}{\partial v^{ij} \partial v_{ij}}.
\ee

Thus we conclude that the two $N{=}4$ supermultiplets \p{V} and \p{U} span a new
nonlinear $N{=}4$ supermultiplet with four physical bosonic and
four fermionic degrees of freedom if they are related as in \p{con1} with the functions
$f$ and $a^{ij}$ obeying to \p{c1},\p{c2}.

It is a rather complicated task to find the general solution of the system \p{c1}, \p{c2}.
Therefore it is desirable to provide some
clarifying examples of systems which could be described with a new nonlinear supermultiplet.
Here we present two of the simplest examples.

\subsection{Hypermultiplet}
It is evident that the simplest solution of the system \p{c1},\p{c2} is given by
\be\label{sol1}
f=1, \qquad a^{ij}=0.
\ee
In this case the resulting $N{=}4$ supermultiplet is the well-known linear hypermultiplet \cite{GR,HP,il1,nlchiral} and the bosonic part of
the action reads
\be\label{sa}
S=2\int dt \left( \frac{\partial^2 L}{\partial u^2}+
2 \frac{\partial^2 L}{\partial v^{ij} \partial v_{ij}}\right) \left[
 \frac{1}{2} {\dot v^{ij}}{\dot v_{ij}}+ {\dot u}^2 \right].
\ee
Clearly, the action \p{sa} describes conformally flat four-dimensional bosonic
manifolds.

\subsection{Nonlinear hypermultiplet and hyper-K\"ahler sigma model}
A more involved example corresponds to the case where both
functions $f$ and $a^{ij}$ depend only on tensor supermultiplet
$v^{ij}$. In this case the equations \p{c1},\p{c2} are simplified
to be \be\label{hk} \frac{\partial a^{ij}}{\partial v^{ij}}=0 ,
\qquad 2\frac{\partial f}{\partial v^{ij}} - \left( \frac{\partial
a^k_i}{\partial v^{kj}}+\frac{\partial a^k_j}{\partial v^{ki}}
\right)=0. \ee As a consequence of \p{hk} the function $f$ has to
be a harmonic one \be \frac{\partial^2}{\partial v^{ij} \partial
v_{ij}} f =0. \ee If we additionally choose the metric $G$
\p{metric1} as
$$G=f$$
then the bosonic part of the action acquires the form
\be\label{sb}
S=\int dt \left[
 \frac{f}{2} {\dot v^{ij}}{\dot v_{ij}}+ \frac{1}{f}\left({\dot u}+
\frac{1}{2}a_{ij}{\dot v^{ij}}\right)^2 \right]. \ee In the action
\p{sb} one may immediately recognize the Gibbons-Hawking Ansatz
for the four-dimensional Hyper-K\"ahler sigma model action with
translational (or triholomorphic) isometry \cite{hk1}, provided
the equations \p{hk} are satisfied. Thus, the $N{=}4$
supersymmetric sigma models with HK geometry in the bosonic sector
may be naturally described within the constructed nonlinear
supermultiplet.

Let us notice that the $N{=}4$ supersymmetric system with the
bosonic action \p{sb} has been firstly constructed in \cite{ks1}
in components. Until now the superfield formulation of the
corresponding nonlinear hypermultiplet has been known only in the
harmonic superspace \cite{delduc1, delduc2}. The constraints
\p{con1} together with the equations \p{hk} provide the superfield
description of the nonlinear hypermultiplet in the standard
$N{=}4$ superspace.

\section{4=2+2}
Another possibility to construct a nonlinear  supermultiplet with
four physical bosons is to start with two chiral $N{=}4$
supermultiplets both containing two physical bosons and then again
identify the fermions in both supermultiplets.

Let us introduce two $N{=}4$ chiral supermultiplets $x$ and $y$ subjected to
ordinary constraints
\be\label{eq1}
D^i x=\bD{}^i \bx =0, \qquad D^i y=\bD{}^i \by =0.
\ee
The most general variant of identification of the fermions in both supermultiplets
reads
\be\label{eq2}
D^i \bx = f D^i \by + g \bD{}^i y, \;
\bD_i x = {\bar f} \bD_i y - {\bar g} D_i \by,
\ee
where the arbitrary functions $f,g$ depend on all superfields $x,\bx,y$ and
$\by$.

The self-consistency of the constraints \p{eq2}  imposes the
following restriction on the functions $f,g$: \be\label{eq3}
g\frac{\partial f}{\partial \bx}= f\frac{\partial g}{\partial
\bx}+\frac{\partial g}{\partial\by}, \qquad {\bar g}\frac{\partial
g}{\partial x}= -{\bar f}\frac{\partial f}{\partial x}-
\frac{\partial f}{\partial y} \ee and their conjugated. It also
follows from \p{eq2} that the auxiliary components of the
superfields $x,y$ \be\label{eq4} A=-\frac{i}{4} \bD{}^2 x|,\;{\bar
A}=-\frac{i}{4}D^2 \bx|, \qquad B=-\frac{i}{4} \bD{}^2 y|,\;{\bar
B}=-\frac{i}{4}D^2 \by| \ee are expressed in terms of physical
bosons and fermions as \be A=\frac{(f{\bar f}+g{\bar g}){\dot
\by}-{\bar f}\;{\dot\bx}}{g}+fermions, \qquad
B=\frac{f{\dot\by}-{\dot\bx}}{g}+fermions. \ee

Now we may construct the most general sigma-model action for this
supermultiplet \be\label{eq5} S=-\frac{1}{16}\int d^4\theta dt
{\cal L}(y,\by, x,\bx). \ee After passing to components, the
bosonic part of the action \p{eq5} reads \be S_B= \int dt G \left[
{\dot x}{\dot{\bar x}}+
 (f{\bar f}+g{\bar g}) {\dot y}{\dot\by} -f{\dot x}{\dot\by}-
{\bar f}{\dot{\bar x}}{\dot y} \right],
\ee
where
\be\label{eq6}
G=\frac{1}{g{\bar g}}\left[ L_{y\by}+{\bar f}L_{x\by}+fL_{{\bar x}y}+
f{\bar f}L_{x{\bar x}}\right].
\ee
The full analysis of this system is out of the scope of the present paper and will be done
elsewhere.

\section{Conclusion}
In this paper we proposed a unified framework for a description of all linear
and nonlinear one-dimensional supermultiplets with $N{=}4$ supersymmetry,
based on ``gluing'' a pair of linear supermultiplets along their fermions.
The functions defining such a gluing obey a system of equations, each solution
of which yields an irreducible supermultiplet, linear or nonlinear.
A given supermultiplet may appear in several equivalent ways which are related
by superfield redefinitions.
It is amazing that {\it all known\/} $N{=}4$ supermultiplets appear in this way,
as we showed explicitly.

Furthermore, by iterating this method, all known $N{=}4$ supermultiplets may
be constructed just from the linear supermultiplet, which features a single
physical boson. Gluing this fundamental ingredient to any other
$N{=}4$ supermultiplet increases $n$ by one, and so any case is eventually
reached starting from several copies of the linear supermultiplet.
In this respect, this supermultiplet plays a role analogous to
the one of the ``root'' supermultiplet~\cite{JG,delduc2,bkmo,delduc3},
which contains four physical bosons (for $N{=}4$ supersymmetry).
However, the root supermultiplet is not unique. There is an infinite
number of supermultiplets with four bosonic and four fermionic components,
while the supermultiplet with one physical boson is unique. Therefore,
we believe that our approach is more general.

Why did we ignore the $n{=}0$ supermultiplets, which do exist in $d{=}1$
supersymmetry? The answer is that gluing such a supermultiplet to an arbitrary
one just expresses the components of the combined multiplet through those of
the arbitrary supermultiplet only. In other words, we merely generate
a superfield redefinition on the arbitrary supermultiplet.

Clearly enough, before discussing the classification issue we should comment
on the completeness of the proposed scheme.
We expect that the general solutions to the equations in Sections 3, 4 and~5
will provide us with novel supermultiplets.
Therefore, one must firstly strive to solve in general the systems of
differential equations presented here.
Secondly, one has to analyze the equivalence relations among all solutions
and characterize the equivalence classes, which are in one-to-one
correspondence with the different supermultiplets.
Thirdly, not all nonlinear $N{=}4$ supermultiplets may be reached directly
by gluing two linear multiplets, so one should investigate the gluing process
with previously found nonlinear multiplets, as seems natural in an iteration,
and also the simultaneous gluing of more than two multiplets.
The associativity of iterated gluing is another issue of interest.
The idea we utilized in this paper may be easily applied to these cases
without any modification. It is only that the ensuing equations are more
involved, and the task of solving them is deferred to future work.

Hence, we only stand at the beginning of a classification program.
Curiously, our framework offers much more information than we looked for.
Indeed, all known supermultiplets correspond merely to the simplest
solutions of our equation systems, e.g.~solutions with a frozen dependence
on one coordinate.
Hence, still open is the most intriguing question:
To which supermultiplets correspond the general solutions?

Finally, we note that the more complicated problem of constructing
$N{=}8$ supermultiplets in one spacetime dimension may also be attacked by
adapting our framework. In this case one has a larger number of possibilities
for gluing together different supermultiplets, but the main needed ingredient
-- the supermultiplet with one physical boson -- is well known.
Moreover, the simplest case of joining two such supermultiplets shall give
birth to a new nonlinear $N{=}8$ supermultiplet with two physical bosons.
We intend to report these results elsewhere.

\section*{Acknowledgements}

S.K. would like to thank the Institute for Theoretical Physics in Hannover,
where this work was completed, for the warm hospitality.

This work has been supported in part by the European Community Human Potential
Program under contract MRTN-CT-2004-005104
\textit{``Constituents, fundamental forces and symmetries of the universe''},
by grants RFBR-06-02-16684, 06-01-00627-a, DFG~436 Rus~113/669/03 and by
INTAS under contract 05-7928.

\end{document}